\documentclass[12pt]{iopart}

\usepackage{iopams}
\usepackage{graphicx,epsfig}
\begin{document}
\hyphenation{phe-no-me-na  pro-ba-bi-li-ty quan-tum le-vels
cor-res-pon-ding e-ner-gy e-ner-gies bar-rier mo-de-ra-te re-le-vant
re-so-nant fol-lo-wing sig-ni-fi-cant va-lues si-mu-la-tions}

\title{Resonance phenomena in Macroscopic Quantum Tunneling: the small
viscosity limit}
\author{Yu N Ovchinnikov}
\address{L.D. Landau Institute for Theoretical Physics, Academy
of Science, Moscow, Russia}
\author{S Rombetto \footnote{also at: Istituto di Cibernetica ``E. Caianiello" del CNR, Pozzuoli (Naples),
Italy}}
\address{Universit\`{a} degli Studi di Napoli ``Federico II", Naples, Italy}
 \author{B Ruggiero}
 \address{Istituto di Cibernetica ``E. Caianiello" del CNR, Pozzuoli (Naples), Italy}
\author{V Corato}
\address{Dipartimento di Ingegneria dell'Informazione, Seconda Universit\`{a}  di Napoli, Naples, Italy}
\author{P Silvestrini}\ead{paolo.silvestrini@unina2.it}
\address{Dipartimento di Ingegneria dell'Informazione, Seconda
Universit\`{a}  di Napoli, Naples, Italy}
\date{\today}
\begin{abstract}
We present a new theoretical approach to describe the quantum
behavior of a macroscopic system interacting with an external
irradiation field, close to the resonant condition. Here we consider
the extremely underdamped regime for a system described by a double
well potential. The theory includes both: transitions from one well
to the other and relaxation processes. We simulate resonant
phenomena in a rf-SQUID, whose parameters lie in the range typically
used in the experiments.  The dependence of the transition
probability W on the external drive of the system $\varphi_x$ shows
three resonance peaks. One   peak is connected with the resonant
tunneling and the two others with the resonant pumping. The relative
position of the two peaks correlated to the resonant pumping depends
on the pumping frequency $\nu$ and on the system parameters. The
preliminary measurements on our devices show  a low dissipation
level and assure that they are good candidates in order to realize
new experiments on the resonant phenomena in the presence of an
external microwave irradiation of proper frequency.
\end{abstract}
\maketitle
\section{Introduction}
\begin{figure}[!t]
\includegraphics[bb = 0 0 216 170]{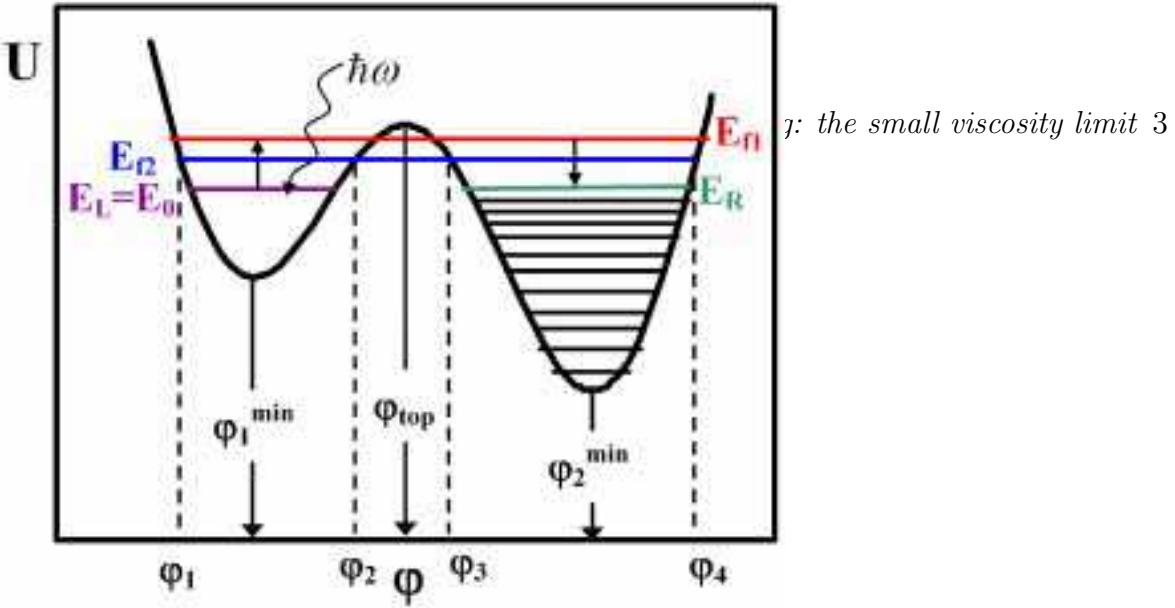}
\caption{\label{potenziale}A sketch of an rf-SQUID double-well
potential U($\varphi$), with two minima located at $\varphi_1^{min}$
and $\varphi_2^{min}$. $\varphi_{top}$ is the position of the
potential maximum. The coordinates
$\varphi_1,\varphi_2,\varphi_3,\varphi_4$ are the ``turning points"
of energy $E_{f_1}$. In the same way it's possible to define turning
points for energies $E_{f_2}$, $E_{0}$, $E_{R}$, $E_{L}$. We study
the case $E_{L}=E_{0}$, that is the ground state in the left
potential well.}
\end{figure}

\begin{figure}
\includegraphics[bb = 0 0 216 160]{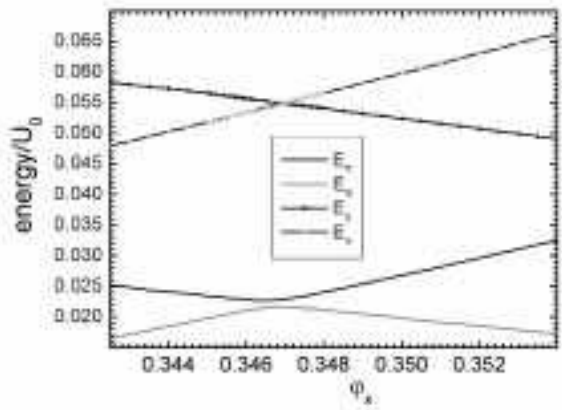} \caption{\label{livelli}The
four essential energy levels considered in this study. All the
quantities are referred to the barrier top. Here $E_{f_{1}}$ and
$E_{f_{2}}$ are the energies of the delocalized levels. $E_0$ is the
first level in the left potential well below $E_{f_2}$ and $E_R$ is
the first level in the right potential well below $E_{f_2}$. }
\end{figure}

Josephson devices are convenient instruments for investigating
macroscopic quantum effects \cite{CaldLegg}. First experiments
focused on incoherent phenomena such as macroscopic quantum
tunneling (MQT) \cite{MQT1, MQT2, luk95, MQT3} and energy level
quantization (ELQ) \cite{ELQ1, ELQ2, ELQ3}. Recent experiments
have shown a coherent superposition of distinct macroscopic
quantum states in  SQUID systems \cite{Lukens, Mooij2,Buisson}
under the action of an external microwave irradiation. These
results have stimulated many researchers to search for new quantum
phenomena at macroscopic level as well as a method to distinguish
them from classical predictions. Moreover quantum effects in SQUID
dynamics are interesting because of possible applications to
quantum computing \cite{LesHouches}, as demonstrated by recent
experiments \cite{Mooij2}, \cite{Buisson}. Anyway these results
need a final confirmation, since some of them have been reproduced
by using a classical model to describe the phenomena
\cite{niels}.\\The main problem to observe quantum phenomena at
macroscopic level is the interaction with the environment which
produces dissipation of coherent effects into classical
 statistical mixture of quantum states. Among a great variety of
 incoherent effects, the environment induces a finite width $\gamma$ of
quantum levels \cite{LO_3} and, in this situation, resonant
phenomena can occur only if $\gamma$ is small compared to the energy
difference between levels, $\hbar\gamma\ll\hbar\Omega_p$, where
\begin{equation}\label{wplasma}
\fl \Omega_p=\sqrt{\frac{1}{M}\left(\frac{\partial^2
 U}{\partial\varphi^2}\right)_{\varphi=\varphi_{min}}}\nonumber
\end{equation}
is the plasma frequency, and $\varphi_{min}$ is the coordinate
corresponding to the minimum of the considered potential well. We
have already discussed \cite{yuri1} the behavior of microwave
induced transitions in a rf-SQUID in the moderate underdamped
regime, when $\gamma$ is larger than the probability of quantum
tunneling under the potential barrier, $T_{N}$, ($\gamma\gg
T_{N}$) \cite{OS}.\\In the present paper we extend the theoretical
analysis to the small viscosity limit, corresponding to
$\gamma\approx T_{N}$, referred as the \emph{extremely underdamped
regime}. In this conditions we can neglect the interwell
transitions with respect to the tunneling transitions. Up to now,
this regime has been investigated only for other Josephson devices
\cite{Mooij2, ELQ3}.  Here we present a new theory specific for
the rf-SQUID, characterized by a double well potential
(Fig.\ref{potenziale}),  in order to identify coherent phenomena
and to suggest a way of
measuring them. \\
It's worth nothing that Averin et al. \cite{Av00}, \cite{Av01},
\cite{LukensPRB} already reported a theoretical investigation of
resonant quantum phenomena in rf SQUIDs. Nevertheless in that paper
all quantities are postulated and not found from the initial
Hamiltonian, while we don't not use any fitting parameters, but only
external parameter values
entering in the Hamiltonian.\\
We study the interaction between a macroscopic quantum system (the
rf-SQUID) and an external microwave irradiation, for frequencies
close to resonant conditions. \\The probability of a transition
under the potential barrier decreases exponentially with the
decreasing energy of the quantum state. Tunneling occurs from a
level close to the barrier top. In this case, as we'll show in the
following, the transition probability curve can present three
peaks.\\A further requirement to observe the quantum tunneling
phenomenon is that the number of levels $n$ is not too large, so
that no crowding of levels happens. At the same time we assume the
number of levels in the potential well is large enough, so that the
motion of a particle with energy close to the top of the potential
barrier can be considered quasiclassical. In this limit we use
approximations
reported in \cite {LO_4}.\\
In the quasiclassical limit, the density of levels close to the
barrier top increases, but in the considered case this effect is not
significant, since numerical factors are of the order of
$\frac{1}{2\pi}\ln n$ \cite{LO_4}. Furthermore, near to the barrier
top, repulsion of levels is relevant and hence the equation for the
system wave functions should be solved exactly. Finally the
tunneling probability of a particle through the potential barrier
depends very strongly on the particle energy E and on the energy
difference between neighboring levels. The position and the width of
each level depends on the external parameters of the system. A
sketch of the double well potential of the rf-SQUID here studied is
reported in Fig.\ref{potenziale}.
\\
Each macroscopic flux state of the rf-SQUID corresponds to the
system wave function confined in one of the two potential wells.
By varying the external parameter $\varphi_x$ it's possible to
control the energy difference between the levels in different
wells. \\Out of resonance conditions, the energy levels are
localized in each well. When  two levels in different wells are in
resonance, a coherent superposition of distinct states  occurs and
a gap appears between the energy levels. As a consequence of the
coherent quantum tunneling process, in exact resonant conditions,
the two levels have an energy tunnel splitting $\Delta$, and the
wave functions spread over the two wells, generating delocalized
states with energies $E_{f_{1}}$ and $E_{f_{2}}$, as shown in
Fig.\ref{livelli}. It's worth noting that the energy tunnel
splitting $\Delta$ corresponds to the tunneling probability $T_N$
in exact resonant conditions and can be experimentally observed
only if resonant levels are very close to
the barrier top and in small viscosity limit \cite{Lukens}.\\
Moreover, to study the tunneling process from the upper levels, it's
necessary to populate them through a pumping process from the ground
state of the left well: this requires the application of a microwave
irradiation of proper frequency $\nu=\omega/2\pi$ and amplitude
${\cal I}$. Even a small amplitude of external microwave irradiation
will cause an essential change in the population of excited states
and will therefore produce a large effect in the tunneling
probability.
\\We study the rf-SQUID dynamics in these particular conditions
and we show some  numerical simulations. For a fixed value of the
pumping frequency $\nu$, the transition probability curve W from
one potential well to the other one (as a function of the external
parameter $\varphi_x$) can present three maxima: one maximum is
connected with the resonant tunneling and the two others with the
resonant pumping toward the two non-localized
levels with energies $E_{f_1}$, $E_{f_2}$.\\
In the considered phenomenon,  friction, that is due to the
interaction with the external environment, has a relevant role
since it leads to a finite width $\gamma$ of levels and destroys
the coherence \cite{LO_7}. In the following sections we will
discuss the induced resonance phenomena by using the density
matrix formalism \cite{matrix}.\\In particular in section II we
discuss the transition probability function in presence of
resonant pumping, in section III we show how to calculate levels
position and transition matrix elements,  in section IV we
determine the energy spectrum in the vicinity of the crossing
point and ,finally, in section V we show the results of numerical
simulations and first experimental data for the dissipation level
for these rf-SQUID based devices.
\section{Transition probability function in presence of  resonant pumping}
In the zero approximation of the system interaction with the thermal
bath, the hamiltonian describing the rf-SQUID in the presence of an
external microwave  with amplitude ${\cal I}$ is
\begin{equation}
\label{potSQUID} \fl  H_0 = -\frac{\hbar^2}{2M}\frac{\partial
^2}{\partial\varphi ^2}+U_0\left[\left(\varphi-\varphi_x\right)^2
+\beta_L\cos\varphi\right]+\frac{{\cal I \hbar}}{2e}\cos(\omega
t)\varphi
\end{equation}
where ${\cal I}$ has the dimensionality of a current, $M$ is the
``mass" of the Josephson junction defined as
\begin{equation}\label{massa}
\fl M=\left(\frac{\hbar}{2e}\right)^2C
\end{equation}
and $C$ is the junction capacitance. Following a common
practice\cite{luk95}, the dimensionless variable $\varphi_x$ is
obtained by referring the magnetic flux to $\Phi_0$/2 and
normalizing it to $\Phi_0$/2$\pi$.
 In eq.({\ref{potSQUID}})
quantities $U_0$ and $\beta_L$ are free parameters of the system
and are defined as
\begin{eqnarray}\label{definizioni}
\fl U_0 =\left[\left(\frac{\Phi_0}{2\pi}\right)^2\frac{1}{L}\right]\\
\fl \beta_L =\frac{2\pi L I_c}{\Phi_0}
\end{eqnarray}
L and $I_c$ are the inductance and the critical current of the
rf-SQUID, respectively. The simplest case is when $U_0$ and
$\beta_L$ are constant, while $\varphi_x$ changes. \\
Our considerations are valid for any potential $U(\varphi)$. The
essential point is that the potential $U(\varphi)$ has the form
given in Fig.\ref{potenziale} in the proximity of the points
$\tilde{\varphi}$. All the parameters $\varphi_x$, $U_0$,
$\beta_L$ depends on the structure of the considered SQUID.
\\Here we
choose parameters such that, near to the top of the barrier, there
are two close levels $E_{f_1},\;E_{f_2}$, that can be considered
as belonging to both potential wells simultaneously (delocalized
levels)
as shown in Fig.\ref{potenziale}. Moreover it's worth noting that
the quantity $\varphi_x$ is proportional to $\frac{IL}{\Phi_0}$.\\
The dissipation is accounted for by an effective resistance
$R_{eff}$ in the RSJ model for the junction \cite{Barone}. So a
bigger $R_{eff}$  implies a better insulation, that means a smaller
width of the levels.  Here we consider an rf-SQUID whose Josephson
junctions presents a large value of resistance $R_{eff}$, so that
the condition
\begin{equation}
\label{disug} \fl \frac{ E_{f_1}-E_{f_2}}{\hbar}\gg\
\gamma_{1},\gamma_{2}
\end{equation}
is satisfied, where  $\gamma_{1},\gamma_{2}$ are the widths of
levels $E_{f_1}$, $E_{f_2}$ respectively. \\To describe
  the resonant tunneling induced by the resonant
pumping quantitatively, we use the following system of equations
\cite{LO_3} for the density matrix elements $\rho^j_f$ {\small
\begin{eqnarray}
\label{densmat1}\fl  \frac{\partial \rho^j_f}{\partial t}=\frac{i{\cal I}}{2e}\cos(\omega t)\sum_{m} \Bigg(  \langle j\vert\varphi\vert m \rangle \exp\left[ -i\left(\frac{E_m-E_j}{\hbar}\right)t \right]\rho_f^m \\
\fl - \langle m\vert\varphi\vert f\rangle \exp\left[
i\left(\frac{E_{m}-E_{f}}{\hbar}\right)t \right]\rho_{m}^{j}\Bigg)
+\sum_{m,n}W_{f\,n}^{j\,m}\rho_{n}^{m} -
\frac{1}{2}\sum_{m}{\left(W_{m\,j}^{m\,j}+W_{m\,f}^{m\,f}\right)\rho_{f}^{j}}\nonumber
\end{eqnarray}
} Here \emph{j, m, f, n} are the considered levels, $E_{j}$,
$E_{m}$, $E_{f}$, $E_{n}$ are their energies and $t$ is the time.\\
The matrix elements $W_{f\,n}^{j\,m}$ entering in
eq.({\ref{densmat1}}) are defined in Appendix A and C. They are
nonvanishing provided that
\begin{equation}
\label{disug2} \fl \vert\left(E_m-E_j \right)-\left(E_n-E_f
\right)\vert \ll\hbar\Omega_p
\end{equation}
If these conditions are satisfied and if the temperature is  low
enough, the nonvanishing terms $W_{f\,n}^{j\,m}$ of the transition
probability matrix are given by the following equation:
\begin{equation} \label{Wmatrix}
\fl
W_{f\,n}^{j\,m}=\frac{\pi}{2R_{eff}e^2}\left(1+\tanh\frac{\tilde{\omega}}{2T}\right)\,N(\tilde{\omega})\cdot
\left[ \langle j\vert e^{\frac{i\varphi}{2}}\vert m\rangle\langle
f\vert e^{-\frac{i\varphi}{2}}\vert n \rangle + \langle j\vert
e^{-\frac{i\varphi}{2}}\vert m\rangle\langle f\vert
e^{\frac{i\varphi}{2}}\vert n \rangle \right]
\end{equation}
where
 \begin{equation}
\label{def_omega} \fl \tilde{\omega}= \frac{\left(E_m-E_j+E_n-E_f
\right)}{\hbar}
\end{equation}
and $N\left({\tilde{\omega}}\right)$ is defined as
\begin{equation}
\label{def_N} \fl N\left({\tilde{\omega}}\right)=
\frac{\tilde{\omega}}{\pi}\coth\left(\frac{\tilde{\omega}}{2T}\right)
\end{equation}
Only the four non-diagonal elements of the density matrix
$\rho_{f_1}^0,\,\rho_{f_2}^0,\,\rho^{f_1}_0,\,\rho^{f_2}_0$ are
nonzero, and they are connected by the expressions
\begin{equation}
\label{def_rho}
 \fl  \rho^{f_1}_0=\left(\rho_{f_1}^0 \right)^*  \hspace{1,2cm}\rho^{f_2}_0=\left(\rho_{f_2}^0 \right)^*
\end{equation}
In presence of an external pumping also the two diagonal elements
$\rho_{f_1}^{f_1}$ and
$\rho_{f_2}^{f_2}$ are non-zero.\\
We suppose that the temperature T is much smaller than the plasma
frequency $\left(k_BT\ll\hbar\Omega_p\right)$, such that it is of
the order  of the energy difference
$k_BT\approx\left(E_{f_1}-E_{f_2}\right)$. In the present article we
consider $T\,=\,50\,mK$. When these hypotheses are satisfied, from
eq.({\ref{densmat1}}) we obtain
\begin{eqnarray}
\label{densmat2} \fl \frac{\partial \rho^0_{f_{1}}}{\partial
t}=-\frac{i{\cal I}}{2e}\cos(\omega t) \langle 0\vert\varphi\vert
f_{1} \rangle e^{\left( -i\frac{E_{f_{1}}-E_0}{\hbar}t
\right)}\rho^0_0+W_{f_{1}\,f_{2}}^{0\,\,0}\rho_{f_{2}}^{0}-\gamma_1\rho_{f_{1}}^{0}\\
\fl \frac{\partial \rho^0_{f_{2}}}{\partial t}=-\frac{i{\cal I}
}{2e}\cos(\omega t) \langle 0\vert\varphi\vert f_{2} \rangle
e^{\left( -i\frac{E_{f_{1}}-E_0}{\hbar}t
\right)}\rho^0_0+W_{f_{2}\,f_{1}}^{0\,\,0}\rho_{f_{1}}^{0}-\gamma_2\rho_{f_{2}}^{0}
\end{eqnarray}
where  the widths $\gamma_1$, $\gamma_2$ of levels are given by the
following expressions
\begin{eqnarray}
\label{def_gamma} \fl
\gamma_1=\frac{1}{2}\left(W_{f_2\,f_1}^{f_2\,f_1}+W_{L\,f_1}^{L\,f_1}+W_{R\,f_1}^{R\,f_1}
\right)\\
\vspace{2mm} \fl
\gamma_2=\frac{1}{2}\left(W_{f_1\,f_2}^{f_1\,f_2}+W_{L\,f_2}^{L\,f_2}+W_{R\,f_2}^{R\,f_2}
\right)\nonumber \end{eqnarray} The indexes L and R refer to the
first energy levels below $E_{f_2}$ in the left and right wells,
respectively.
 The tunneling transition probability
between two states in different wells decreases very quickly  for
states below the barrier top, such as $e^{-2\pi n}$, where $n$ is
the number of states counted from the barrier top \cite{LO_2}. As
a consequence we can neglect transitions into lower levels in the
right potential well. \\In order to study this phenomenon, only
five energy levels ($E_{f_1}$, $E_{f_2}$, $E_0$, $E_L$, $E_R$) are
essential quantities. $E_0$ is the energy of the ground state in
the left potential well, $E_L$ is the first level in the left
potential well below $E_{f_2}$ (in the simulations presented here
$E_L$=$E_0$), $E_R$ is the first level in the right potential well
below $E_{f_2}$. Energy levels as a function of the external
parameter $\varphi_x$, are shown in Fig.\ref{livelli}. The pumping
frequency $\nu$ is supposed to be close to the frequencies
$E_{f_1}/\hbar$ and $E_{f_2}/\hbar$ (referring both to
$E_{0}/\hbar$). In our simulations $E_L$ is also the ground state
of the left potential well (it's the same as $E_{0}$), while in
the right well there are many more levels, not involved in the
described process.
\\If these conditions are satisfied, the solutions of
eqs.({\ref{densmat2}}), (13) are:
%\begin{widetext}
\begin{eqnarray}
\label{densmat3}
\fl \rho^0_{f_1}=A_{f_1}e^{i\left(\omega-\frac{E_{f_1}-E_0}{\hbar}\right)t}+B_{f_1}e^{i\left(\omega-\frac{E_{f_2}-E_0}{\hbar}\right)t}\\
\fl
\rho^0_{f_2}=B_{f_2}e^{i\left(\omega-\frac{E_{f_1}-E_0}{\hbar}\right)t}+A_{f_2}e^{i\left(\omega-\frac{E_{f_2}-E_0}{\hbar}\right)t}\nonumber
\end{eqnarray}
%\end{widetext}
 In eqs.({\ref{densmat3}}) coefficients
$A_{f_1},\,A_{f_2},\,B_{f_1},\,B_{f_2}$ are defined as follows
{\scriptsize
\begin{eqnarray}
\label{def_AB} \fl A_{f_1}=-\frac{{\cal I}}{4e}\langle
0\vert\varphi\vert f_1\rangle\left[\omega-\frac{E_{f_1}-E_0}{\hbar}-i\gamma_1+\frac{W_{f_1\,f_2}^{0\,0}W_{f_2\,f_1}^{0\,0}}{\omega-\frac{E_{f_1}-E_0}{\hbar}-i\gamma_2} \right]^{-1}\nonumber\\
\fl A_{f_2}=-\frac{{\cal I}}{4e}\langle
0\vert\varphi\vert f_2\rangle\left[\omega-\frac{E_{f_2}-E_0}{\hbar}-i\gamma_2+\frac{W_{f_1\,f_2}^{0\,0}W_{f_2\,f_1}^{0\,0}}{\omega-\frac{E_{f_2}-E_0}{\hbar}-i\gamma_1} \right]^{-1}\\
\fl B_{f_1}=-\frac{iW_{f_1\,f_2}^{0\,0}A_{f_2}}{\omega-\frac{E_{f_2}-E_0}{\hbar}-i\gamma_1}\nonumber\\
\fl
B_{f_2}=-\frac{iW_{f_2\,f_1}^{0\,0}A_{f_1}}{\omega-\frac{E_{f_1}-E_0}{\hbar}-i\gamma_2}\nonumber
\end{eqnarray}}
The diagonal elements $\rho_{f_1}^{f_1}$ and $\rho_{f_2}^{f_2}$ of
the density matrix can be find by solving the following equations
\begin{eqnarray}
\label{rho_systa} \fl \frac{\partial \rho^{f_1}_{f_1}}{\partial
t}=\frac{i{\cal I}}{2e}\cos(\omega t) \left[\rho^{0}_{f_1}\langle
f_1\vert\varphi\vert 0 \rangle e^{i\left(\frac{E_{f_1}-E_0}{\hbar}
\right)t}-\rho^{f_1}_{0}\langle 0\vert\varphi\vert f_1 \rangle
e^{\left(-i\frac{E_{f_1}-E_0}{\hbar}t\right)}\right]
+W_{f_1\,f_2}^{f_1\,f_2}\rho_{f_2}^{f_2}-2\gamma_1\rho_{f_1}^{f_1}\\
\label{rho_systb} \fl \frac{\partial \rho^{f_2}_{f_2}}{\partial t}
=\frac{i{\cal I}}{2e}\cos(\omega t)\left[\rho^{0}_{f_2}\langle
f_2\vert\varphi\vert 0 \rangle e^{i
\left(\frac{E_{f_2}-E_0}{\hbar}\right)t}-\rho^{f_2}_{0}\langle
0\vert\varphi\vert f_2 \rangle e^{-i
\left(\frac{E_{f_2}-E_0}{\hbar}\right)t }\right]
+W_{f_2\,f_1}^{f_2\,f_1}\rho_{f_1}^{f_1}-2\gamma_2\rho_{f_2}^{f_2}
\end{eqnarray}
These rate equations completely describe the dynamics of the
process. The solutions of this system are respectively:
\begin{eqnarray}
\label{rho_syst1a} \fl
\rho_{f_1}^{f_1}\left(t\right)=\hat{\rho}_{f_1}^{f_1}+{\cal F}_1
e^{i \left(\frac{E_{f_2}-E_{f_1}}{\hbar}\right)t} +{\cal
F}_1^{*}e^{-i \left(\frac{E_{f_2}-E_{f_1}}{\hbar}\right)t}\\
\fl \label{rho_syst1b}
\rho_{f_2}^{f_2}\left(t\right)=\hat{\rho}_{f_2}^{f_2}+{\cal D}_1
e^{i \left(\frac{E_{f_2}-E_{f_1}}{\hbar}\right)t}+{\cal
D}_1^{*}e^{-i \left(\frac{E_{f_2}-E_{f_1}}{\hbar}\right)t}
\end{eqnarray}
The expressions for ${\cal F}_1$ and ${\cal D}_1$ are given in
Appendix A. By using equations ({\ref{rho_syst1a}}) and
({\ref{rho_syst1b}}) inside equations ({\ref{rho_systa}}) and
({\ref{rho_systb}}), we obtain the expressions of the quantities
$\hat{\rho}_{f_1}^{f_1}$ and $\hat{\rho}_{f_2}^{f_2}$  as following:
\begin{eqnarray}
\label{rho_syst2a}\fl  \hat{\rho}_{f_1}^{f_1} = \frac{{\cal
I}^2}{16e^2}\frac{1}{\gamma_1\gamma_2-\frac{1}{4}W_{f_2\,f_1}^{f_2\,f_1}W_{f_1\,f_2}^{f_1\,f_2}}\nonumber\\
\fl \cdot\Bigg[\gamma_2\mid\langle 0\vert\varphi\vert f_1
\rangle\mid^2\frac{\gamma_1-\gamma_2W_{f_1\,f_2}^{0\,0}W_{f_2\,f_1}^{0\,0}\left[\left(\omega-\frac{E_{f_1}-E_{0}}{\hbar}
\right)^2+\gamma_2^2\right]^{-1}}{\left(\omega-\frac{E_{f_1}-E_{0}}{\hbar}\right)^2+\gamma_1^2+\frac{\left(W_{f_1\,f_2}^{0\,0}W_{f_2\,f_1}^{0\,0}\right)^2+2W_{f_1\,f_2}^{0\,0}W_{f_2\,f_1}^{0\,0}\left[\left(\omega-\frac{E_{f_1}-E_{0}}{\hbar}
\right)^2-\gamma_1\gamma_2\right]}{\left(\omega-\frac{E_{f_1}-E_{0}}{\hbar}
\right)^2+\gamma_2^2} }\nonumber\\
\fl +\frac{1}{2}W_{f_1\,f_2}^{f_1\,f_2}\mid\langle
0\vert\varphi\vert
f_2\rangle\mid^2\frac{\gamma_2-\gamma_1W_{f_1\,f_2}^{0\,0}W_{f_2\,f_1}^{0\,0}\left[\left(\omega-\frac{E_{f_2}-E_{0}}{\hbar}\right)^2+\gamma_1^2\right]^{-1}}{\left(\omega-\frac{E_{f_2}-E_{0}}{\hbar}\right)^2+\gamma_2^2+\frac{\left(W_{f_1\,f_2}^{0\,0}W_{f_2\,f_1}^{0\,0}\right)^2+2W_{f_1\,f_2}^{0\,0}W_{f_2\,f_1}^{0\,0}\left[\left(\omega-\frac{E_{f_2}-E_{0}}{\hbar}\right)^2-\gamma_1\gamma_2\right]}{\left(\omega-\frac{E_{f_2}-E_{0}}{\hbar}\right)^2+\gamma_1^2}}\Bigg]
\end{eqnarray}
\begin{eqnarray}
\label{rho_syst2b}\fl  \hat{\rho}_{f_2}^{f_2}=\frac{{\cal
I}^2}{16e^2}\frac{1}{\gamma_1\gamma_2-\frac{1}{4}W_{f_2\,f_1}^{f_2\,f_1}W_{f_1\,f_2}^{f_1\,f_2}}\nonumber\\
\fl \cdot\Bigg[\gamma_1\mid\langle 0\vert\varphi\vert f_2
\rangle\mid^2\frac{\gamma_2-\gamma_1W_{f_2\,f_1}^{0\,0}W_{f_1\,f_2}^{0\,0}\left[\left(\omega-\frac{E_{f_2}-E_{0}}{\hbar}
\right)^2+\gamma_1^2 \right]^{-1}
}{\left(\omega-\frac{E_{f_2}-E_{0}}{\hbar}\right)^2+\gamma_2^2
+\frac{\left(W_{f_1\,f_2}^{0\,0}W_{f_2\,f_1}^{0\,0}\right)^2 +
2W_{f_1\,f_2}^{0\,0}W_{f_2\,f_1}^{0\,0}\left[\left(\omega-\frac{E_{f_2}-E_{0}}{\hbar}
\right)^2-\gamma_1\gamma_2
\right]}{\left(\omega-\frac{E_{f_2}-E_{0}}{\hbar}
\right)^2+\gamma_1^2} } \nonumber\\
\fl +\frac{1}{2}W_{f_2\,f_1}^{f_2\,f_1}\mid\langle
0\vert\varphi\vert f_1 \rangle\mid^2
\frac{\gamma_{1}-\gamma_{2}W_{f_1\,f_2}^{0\,0}W_{f_2\,f_1}^{0\,0}\left[\left(\omega-\frac{E_{f_1}-E_{0}}{\hbar}
\right)^2+\gamma_2^2 \right]^{-1}}
{\left(\omega-\frac{E_{f_1}-E_{0}}{\hbar}\right)^2+\gamma_1^2+
 {\frac{\left(W_{f_1\,f_2}^{0\,0}W_{f_2\,f_1}^{0\,0}\right)^2+
2W_{f_1\,f_2}^{0\,0}W_{f_2\,f_1}^{0\,0}\left[\left(\omega-\frac{E_{f_1}-E_{0}}{\hbar}
\right)^2-\gamma_1\gamma_2\right]}{\left(\omega-\frac{E_{f_1}-E_{0}}{\hbar}\right)^2+\gamma_2^2}}}\Bigg]
\end{eqnarray}
Finally we can introduce the transition probability $W$ from the
left to the right potential well. $W$ is a function of the
external parameter $\varphi_x$ and computing this quantity is the
main goal of the paper. The transition probability is given by the
expression
\begin{equation}
\label{Wgenerica}
 \fl W=W_{R\,f_1}^{R\,f_1}\rho_{f_1}^{f_1}+W_{R\,f_2}^{R\,f_2}\rho_{f_2}^{f_2}
\end{equation}
where, in the low temperature limit, here considered, from
eq.(\ref{Wmatrix}) we know that
\begin{eqnarray}
\label{WRF1_2} \fl
W_{R\,f_1}^{R\,f_1}=\frac{2\left(E_{f_1}-E_{R}\right)}{R_{eff}e^2}\mid\langle
R\mid e^{i \varphi/2}\mid f_1\rangle\mid ^2\\
\fl
W_{R\,f_2}^{R\,f_2}=\frac{2\left(E_{f_2}-E_{R}\right)}{R_{eff}e^2}\mid\langle
R\mid e^{i \varphi/2}\mid f_2\rangle\mid ^2
\end{eqnarray}
while the expressions for $\rho_{f_1}^{f_1}$ and $\rho_{f_2}^{f_2}$
are given in eqs.(\ref{rho_syst1a}) and (\ref{rho_syst1b})
respectively. In the  eqs. (\ref{rho_syst2a}) and
(\ref{rho_syst2b}), the quantities ${\cal F}_1$ and ${\cal D}_1$
determinate the value of the oscillating part of the tunneling
probability $W$. \\It's worth noting that change of the external
parameter $\varphi_x$ leads to a redistribution of wave functions
relative to energies $E_{f_1}$, $E_{f_2}$ between left and right
potential wells, so that transition matrix elements  and widths of
levels $\gamma_{1}$, $\gamma_{2}$ are functions of the external
parameter $\varphi_x$. To complete our considerations we will find
the levels position and the transition matrix elements as a function
of the
external parameters, as shown in the next section.\\
\section{Levels position and transition matrix elements.}
As we have seen before, a necessary condition for resolving the
splitting is that the experimental linewidth of states
($\gamma_{1}$, $\gamma_{2}$) be smaller than $\Delta$
$(\Delta\gg\hbar\gamma_{1}$, $\hbar\gamma_{2})$: it
 can be satisfied only
for levels near to the barrier top (Fig.\ref{potenziale}). In
Fig.\ref{potenziale} $\varphi_1,\varphi_2,\varphi_3,\varphi_4$ are
the ``turning points", that are the solutions of equation
\begin{equation}\label{energy}
 \fl U(\varphi_{1,2,3,4})=E
\end{equation}
where $E$ is the energy value. In Fig.\ref{potenziale} the points
$\varphi_{1}^{min}$, $\varphi_{2}^{min}$ are coordinates
corresponding to the minima of potential $U(\varphi)$ and the point
$\varphi_{top}$ is the coordinate corresponding to the maximum of
the potential $U(\varphi)$. For energy values $E$ close to the top
of the barrier, the wave function $\Psi$ can be written in the form
\begin{eqnarray}\label{f_onda}
\fl
\Psi\,=\,G^{-1}\cdot\frac{1}{\left(2M\left(E-U(\varphi)\right)\right)^{1/4}}\sin\left(\frac{\pi}{4}+\int\limits_{\varphi_1}
{\sqrt{2M\left(E-U\left(\varphi\right)\right)}}d\varphi\right) \textrm{\small{in the left potential well}}\nonumber\\
\fl \Psi\,=\,A\cdot {\cal D}_{-\frac{1-i \lambda}{2}}((1+i
)(2MU_1)^{1/4}(\varphi-\varphi_{top}))+A^{*}{\cal
D}_{-\frac{1+i \lambda}{2}}\left((1-i )(2MU_1)^{1/4}(\varphi-\varphi_{top})\right) \textrm{\small{near the barrier top}}\\
\fl
\Psi\,=\,\frac{B}{\left(2M\left(E-U(\varphi)\right)\right)^{1/4}}\sin\left(\frac{\pi}{4}+\int\limits^{\varphi_4}
{\sqrt{2M\left(E-U(\varphi)\right)}}d\varphi\right)
\textrm{\small{in the right potential well}}\nonumber
\end{eqnarray}
In eqs.(\ref{f_onda}) ${\cal D}_{p}$ is the parabolic cylinder
function \cite{integrali}, while quantities A and B are numerical
factors, G is a normalization factor, defined in the following. The
other quantities are defined as
\begin{eqnarray}\label{def_U}
\fl U_{top}=U(\varphi_{top})\\\label{def_U_1} \fl
U_1=-\frac{1}{2}\left(\frac{\partial^2
U}{\partial\varphi^2}\right)_{\varphi=\varphi_{top}}\\
\label{def_lambda} \fl \lambda=\sqrt{2MU_{1}}\frac{U_{top}-E}{U_{1}}
\end{eqnarray}
First expression in eqs.(\ref{f_onda}) is valid in the left
potential well, the second one is valid close to the barrier top,
the third one is valid in the right potential well. Coefficients A,
B are defined by imposing boundary conditions for these three
expressions. Near to the top of the potential barrier we use the
exact solutions of the Schroedinger equation for the considered
system, whereas in the vicinity of the points $\varphi_1$,
$\varphi_4$ we use the well note method of outgoing in complex
plane\cite{LL}. The relation between coefficients A and B can be
found by matching coefficients in the right potential well
\begin{equation}
\label{defA_B}
 \fl A\,=\,\frac{B}{2^{3/4}(2MU_1)^{1/8}}\exp\left[\frac{\pi\lambda}{8}-\frac{i
\pi}{8}+\frac{i \lambda}{4}+\frac{i
\lambda}{4}\ln\left(\frac{2}{\lambda}\right)+i
\int\limits_{\varphi_3}^{\varphi_4}{d\varphi\sqrt{2M\left(E-U(\varphi)\right)}}
\right]
\end{equation}
 By matching boundary  conditions in the left potential we obtain
two equations. The first one is the exact equation for the position
of the levels near the barrier top in the case of small viscosity
limit {\small
\begin{eqnarray}\label{spectrum}
\fl
\cos\left(\int\limits_{\varphi_3}^{\varphi_4}{d\varphi\sqrt{2M\left(E-U(\varphi)\right)}}-\int\limits_{\varphi_1}^{\varphi_2}{d\varphi\sqrt{2M\left(E-U(\varphi)\right)}}\right)+
\left(1+\exp(-\pi\lambda)\right)^{1/2}\cdot\nonumber\\
\fl
\cdot\cos\left(\chi+\frac{\lambda}{2}+\frac{\lambda}{2}\ln\left(\frac{2}{\lambda}\right)+
\int\limits_{\varphi_3}^{\varphi_4}{d\varphi\sqrt{2M\left(E-U(\varphi)\right)}}+\int\limits_{\varphi_1}^{\varphi_2}{d\varphi\sqrt{2M\left(E-U(\varphi)\right)}}
\right)=0
\end{eqnarray}
} In eq.(\ref{spectrum}) phase shift $\chi$ is defined by the
expression
\begin{equation}\label{Gamma}
\fl \Gamma\left(\frac{1+i \lambda}{2}
\right)\,=\,\frac{\sqrt{2\pi}e^{(-\pi\lambda
/4)}}{\sqrt{1+e^{(-\pi\lambda)}}}e^{\left(i \chi\right)}
\end{equation}
where $\Gamma(x)$ is the Euler gamma function and the phase $\chi$
can be presented in the form
\begin{equation}\label{eq_chi}
\fl \chi = \frac{\lambda}{2}\Psi
\left(\frac{1}{2}\right)-\sum_{k=0}^{\infty}\left(\arctan\left(\frac{\lambda}{2k+1}
\right)-\frac{\lambda}{2k+1}\right)
\end{equation}
In eq.(\ref{eq_chi}) $\psi(x)$ is the Euler function and
\begin{equation*}
\fl \psi\left(\frac{1}{2}\right)=-C-2\ln 2\,=\, -1.96351
\end{equation*}
The second equation gives the value of the coefficient B as follows
\begin{eqnarray}\label{def_B}
\fl
B=\exp\left(\frac{\pi\lambda}{2}\right)\Bigg[\left(1+\exp\left(-\pi\lambda\right)\right)^{1/2}\sin\left(\chi+\frac{\lambda}{2}+\frac{\lambda}{2}\ln\left(\frac{2}{\lambda}\right)
+\int\limits_{\varphi_1}^{\varphi_2}{d\varphi\sqrt{2M\left(E-U(\varphi)\right)}}+\int\limits_{\varphi_3}^{\varphi_4}{d\varphi\sqrt{2M\left(E-U(\varphi)\right)}}\right)\nonumber\\
\fl
-\sin\left(\int\limits_{\varphi_1}^{\varphi_2}{d\varphi\sqrt{2M\left(E-U(\varphi)\right)}}-\int\limits_{\varphi_3}^{\varphi_4}{d\varphi\sqrt{2M\left(E-U(\varphi)\right)}}\right)\Bigg]
\end{eqnarray}
In order to obtain  two close levels near to the barrier top (but
not too close) the parameter $\lambda$ should be $\lambda\geq 1$. In
such a case, the normalization factor G can be approximated as
\begin{equation}\label{def_G}
\fl G^2 =
\frac{\hbar}{2}\left(\int\limits_{\varphi_1}^{\varphi_2}\frac{d\varphi}{\sqrt{2M\left(E-U(\varphi)\right)}}+B^2\int\limits_{\varphi_3}^{\varphi_4}\frac{d\varphi}{\sqrt{2M\left(E-U(\varphi)\right)}}\right)
\end{equation}
Since we consider the energy E close to the barrier top, it's
possible to use  expressions with explicit energy dependence
(perturbation theory over the quantity ($U_{top} - E$) with
separation of the singular terms):
\begin{eqnarray}\label{int_1}
\fl
\int_{\varphi_1}^{\varphi_2}d\varphi\sqrt{2M\left(E-U(\varphi)\right)}=
\int_{\tilde{\varphi_1}}^{\varphi_{top}}d\varphi\sqrt{2M\left(U_{top}-U(\varphi)\right)}-\left(U_{top}-E\right)\sqrt{\frac{M}{2}}\int_{\tilde{\varphi_1}}^{\varphi_{top}}d\varphi\left(\frac{1}
{\sqrt{\left(U_{top}-U(\varphi)\right)}}-\frac{\sqrt{\varphi_{top}-\tilde{\varphi_1}}}{\left(\varphi_{top}-\varphi\right)\sqrt{U_1\left(\varphi
-\tilde{\varphi_1}\right) }} \right)\nonumber\\
\fl
-\left(U_{top}-E\right)\sqrt{\frac{M}{2U_1}}\left[\ln\left(\frac{8\left(\varphi_{top}-\tilde{\varphi_1}\right)\sqrt{U_1}
}{\sqrt{U_{top}-E}}\right)+\frac{1}{2}\right]
\end{eqnarray}
\begin{eqnarray}\label{int_2}
\fl \int_{\varphi_3}^{\varphi_4}
d\varphi\sqrt{2M\left(E-U(\varphi)\right)}=
\int_{\varphi_{top}}^{\tilde{\varphi_4}}d\varphi\sqrt{2M\left(U_{top}-U(\varphi)\right)}
-\left(U_{top}-E\right)\sqrt{\frac{M}{2}}\int_{\varphi_{top}}^{\tilde{\varphi_4}}d\varphi\left(\frac{1}
{\sqrt{\left(U_{top}-U(\varphi)\right)}}
-\frac{\sqrt{\tilde{\varphi_4}-\varphi_{top}}}{\left(\varphi-\varphi_{top}\right)\sqrt{U_1\left(\tilde{\varphi_4}-\varphi
\right) } }  \right)\nonumber\\
\fl
-\left(U_{top}-E\right)\sqrt{\frac{M}{2U_1}}\left[\ln\left(\frac{8\left(\tilde{\varphi_4}-\varphi_{top}\right)\sqrt{U_1}
}{\sqrt{U_{top}-E}}\right)+\frac{1}{2}\right]
\end{eqnarray}
In eqs.(\ref{int_1}), (\ref{int_2}) quantities
$\tilde{\varphi}_{1}$, $\tilde{\varphi}_{4}$ are defined as
\begin{eqnarray*}\label{def_phitilde}
\fl \tilde{\varphi}_{1}=\varphi_{1}\,\left(E=U_{top}\right) \\
\fl \tilde{\varphi}_{4}=\varphi_{4}\,\left(E=U_{top}\right)
\end{eqnarray*} Energies $E_{L}$, $E_{R}$ of states $\Psi_{L}$, $\Psi_{R}$ can be
found by using eqs.(\ref{spectrum}), (\ref{int_1}), (\ref{int_2}),
provided that $\Psi_{L}$, $\Psi_{R}$ are referred to the barrier
top.  In this case wavefunctions $\Psi_{L}$, $\Psi_{R}$ can be
described by using the quasiclassical approximation, so that we find
\begin{eqnarray}\label{def_psiL}
\fl
\Psi_{L}=\frac{1}{G_L}\frac{\sin\left(\frac{\pi}{4}+\int\limits_{\varphi_1}^{\varphi}d\varphi\sqrt{2M\left(E-U(\varphi)
\right) } \right)}{\left(2M\left(E-U(\varphi)\right)\right)^{1/4}}\\
\label{def_psiR} \fl
\Psi_{R}=\frac{1}{G_R}\frac{\sin\left(\frac{\pi}{4}+\int\limits_{\varphi}^{\varphi_4}d\varphi\sqrt{2M\left(E-U(\varphi)
\right) } \right)}{\left(2M\left(E-U(\varphi)\right)\right)^{1/4}}
\end{eqnarray}
where the factors $G_L$ and $G_R$ are defined as
\begin{eqnarray}\label{def_GLR}
\fl
G_L^2=\frac{\hbar}{2}\int\limits_{\varphi_1}^{\varphi_2}{\frac{d\varphi}{\sqrt{2M\left(E-U(\varphi)\right)}}}
\\\fl G_R^2=\frac{\hbar}{2}\int\limits_{\varphi_3}^{\varphi_4}{\frac{d\varphi}{\sqrt{2M\left(E-U(\varphi)\right) }}}
\end{eqnarray}
The wave function $\Psi_0$ of the ground state can be taken in the
form
\begin{equation}\label{psi0}
\fl
\Psi_0=\frac{\left(2MU_1^{min}\right)^{1/8}}{\pi^{1/4}}\;e^{\left(-\int\limits_{\varphi_1^{min}}^{\varphi}d\varphi
\sqrt{2M\left(U(\varphi)-U(\varphi_1^{min})\right) }\right) }
\end{equation}
where $U_{1,2}^{min}=U(\varphi_{1,2}^{min})$ and
$\varphi_{1,2}^{min}$ are the coordinates  corresponding to the
position of the left and right minimum in the potential $U(\varphi)$
\begin{eqnarray}\label{min}
\fl  \left(\frac{\partial U}{\partial\varphi}
\right)_{\varphi =\varphi_{1,2}^{min}}=0\nonumber\\
\fl
U(\varphi)=U(\varphi_{1,2}^{min})+U_{1,2}^{min}(\varphi-\varphi_{1,2}^{min})^2+...\nonumber
\end{eqnarray}
\section{The energy spectrum in the vicinity of the crossing point.}
In order to describe the spectrum near to the crossing point we
introduce two functions $\Phi_1$ and $\Phi_2$, defined as:
\begin{eqnarray}\label{defPhi_1}
\fl
\Phi_{1}=\int\limits_{\varphi_1}^{\varphi_2}{d\varphi\sqrt{2M\left(E-U(\varphi)\right)
}+\frac{\lambda}{4}\left(1+\ln\left(\frac{2}{\lambda}\right)\right)}=
\int\limits_{\tilde{\varphi_1}}^{\varphi_{top}}
{d\varphi\sqrt{{2M\left(U_{top}-U(\varphi)\right)}}}\nonumber\\
\fl
-\left(U_{top}-E\right)\sqrt{\frac{M}{2}}\int_{\tilde{\varphi_1}}^{\varphi_{top}}d\varphi\left(\frac{1}{\sqrt{U_{top}-U(\varphi)}}-
\frac{\sqrt{\varphi_{top}-\tilde{\varphi_1}}}{\left(\varphi_{top}-\varphi\right)\left(\sqrt{U_1\left(\varphi-\tilde{\varphi_1}\right)}\right)}
\right)\nonumber\\
\fl +\left(U_{top}-E\right)\sqrt{\frac{M}{2U_1}}
\ln\left(\frac{2^{1/4}}{8\left(MU_1\right)^{1/4}\left(\varphi_{top}-\tilde{\varphi_1}
\right)} \right)
\end{eqnarray}
\begin{eqnarray}\label{defPhi_2}
\fl
\Phi_2=\int_{\varphi_3}^{\varphi_4}d\varphi\sqrt{2M\left(E-U(\varphi)\right)
}+\frac{\lambda}{4}\left(1+\ln\left(\frac{2}{\lambda}\right)\right)=\int_{\varphi_{top}}^{\tilde{\varphi_4}
} d\varphi\sqrt{{2M\left(U_{top}-U(\varphi)\right)}}\nonumber\\
\fl
-\left(U_{top}-E\right)\sqrt{\frac{M}{2}}\int_{\varphi_{top}}^{\tilde{\varphi_4}}d\varphi\left(\frac{1}{\sqrt{U_{top}-U(\varphi)}}-
\frac{\sqrt{\tilde{\varphi_4}-\varphi_{top}}}{\left(\varphi-\varphi_{top}\right)\left(\sqrt{U_1\left(\tilde{\varphi_4}-\varphi\right)}\right)}
\right)\nonumber\\
\fl +\left(U_{top}-E\right)\sqrt{\frac{M}{2U_1}}
\ln\left(\frac{2^{1/4}}{8\left(MU_1\right)^{1/4}\left(\tilde{\varphi_4}-\varphi_{top}
\right)} \right)
\end{eqnarray}
Now, we suppose that for some value of the external parameters
$\varphi_x^0, U_0, \beta_L$ there is a point characterized by $E_0$,
$\lambda_0=\lambda\left(E_0\right)$ (defined in eq.\ref{def_lambda})
such that
\begin{eqnarray}\label{eqPhi1,2}
\fl  \Phi_1\left(U_0, \beta_L, \varphi_x^0, E_0,
\lambda_0\right)+\frac{1}{2}\chi\left(\lambda_0\right)=\frac{\pi}{2}+\pi
k_1\\
\fl \Phi_2\left(U_0, \beta_L, \varphi_x^0, E_0,
\lambda_0\right)+\frac{1}{2}\chi\left(\lambda_0\right)=\frac{\pi}{2}+\pi
k_2
\end{eqnarray}
where $k_{1}$, $k_{2}$ are integer numbers. The equations (46),(47)
determine the "crossing point". In fact, we are interested to find
the external parameter $\varphi_x$ such that the two energy levels
$E_{f1}$ and $E_{f2}$ are close. As a consequence, next to this
special point, we can use Taylor expansion for the functions
$\Phi_1$ and $\Phi_2$ and we can write:
\begin{eqnarray}\label{eqPhi1}
\fl  \Phi_1 +\frac{1}{2}\chi=\frac{\pi}{2}+\pi
k_1+\alpha_1\delta\varphi_x + \beta_1\delta E\\ \label{eqPhi2b}\fl
\Phi_2+\frac{1}{2}\chi=\frac{\pi}{2}+\pi k_2+\alpha_2\delta\varphi_x
+ \beta_2\delta E
\end{eqnarray}
where
\begin{equation}
\fl E=E_{0}+\delta E, \;\;\;\;
\varphi_x=\varphi_x^{0}+\delta\varphi_x
\end{equation}
and the quantities $\alpha_{1}$, $\alpha_{2}$, $\beta_{1}$,
$\beta_{2}$ can be found from eqs.(\ref{potSQUID}),
(\ref{defPhi_1}). They are given in Appendix B. From
eqs.(\ref{spectrum}), (\ref{eqPhi1}), (\ref{eqPhi2b}) we obtain the
equation to calculate the energy spectrum for two close levels near
to the barrier top
\begin{eqnarray}\label{deltaE}
\fl  \beta_1\beta_2(\delta
E)^2+\alpha_1\alpha_2(\delta\varphi_x)^2\nonumber
\\ \fl +(\alpha_1\beta_2+\alpha_2\beta_1)\delta\varphi_x\delta E -
\frac{1}{4}e^{(-\pi\lambda)}=0
\end{eqnarray}
The solutions of this equation are two hyperboles
\begin{equation}\label{hyperboles}
\fl \delta
E=-\frac{1}{2\beta_1\beta_2}\left[(\alpha_1\beta_2+\alpha_2\beta_1)\delta\varphi_x\pm\sqrt{(\alpha_1\beta_2-\alpha_2\beta_1)^2(\delta\varphi_x)^2+\beta_1\beta_2\exp(-\pi\lambda)}\right]
\end{equation}
Note that Eq.\ref{hyperboles} gives the position of two close levels
as function of the external parameter $\varphi_x$. Two levels can be
defined close if $\delta E$ $<<$ $\hbar\Omega_p$ (the plasma
frequency, defined in the introduction) and this occurs if the
parameter $\exp(-\pi\lambda)$ is small, that is if $\lambda$ is of
the order of one.
\begin{figure}[h]
\includegraphics[bb =  0 0 216 169]{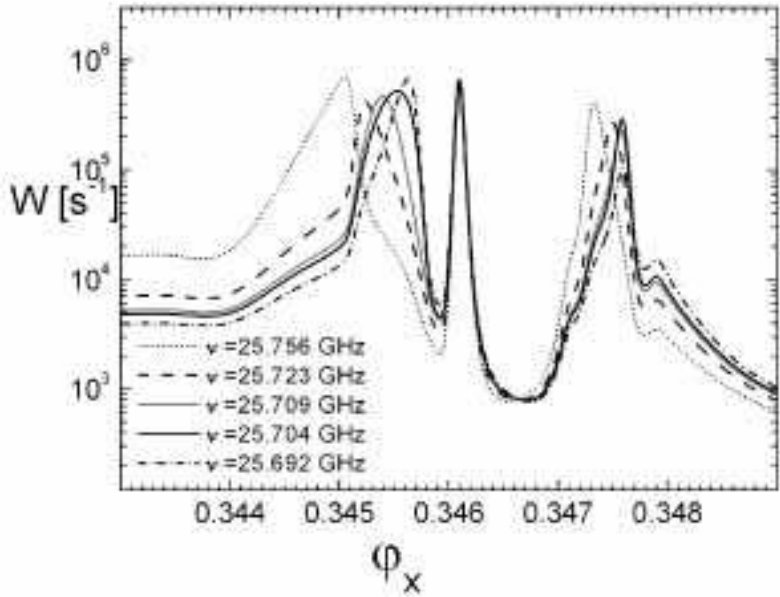}
\caption{\label{Wtre}Transition probability
 W vs. $\varphi_x$ for different values of the pumping
frequency $\nu$. W presents three peaks: the central one is due to
the resonance tunneling and two lateral ones are due to the resonant
pumping of the two levels close to the barrier top. Curves are
obtained by using the following parameters for the numerical
simulation: $\beta_L$=1.75, L=210 pH, C=0.1 pF, and $R_{eff}$=8
M$\Omega$.}
\end{figure}\\
\section{Numerical simulations and preliminary measurements}
\begin{figure}[!h]
\includegraphics[bb = 0 0 216 168]{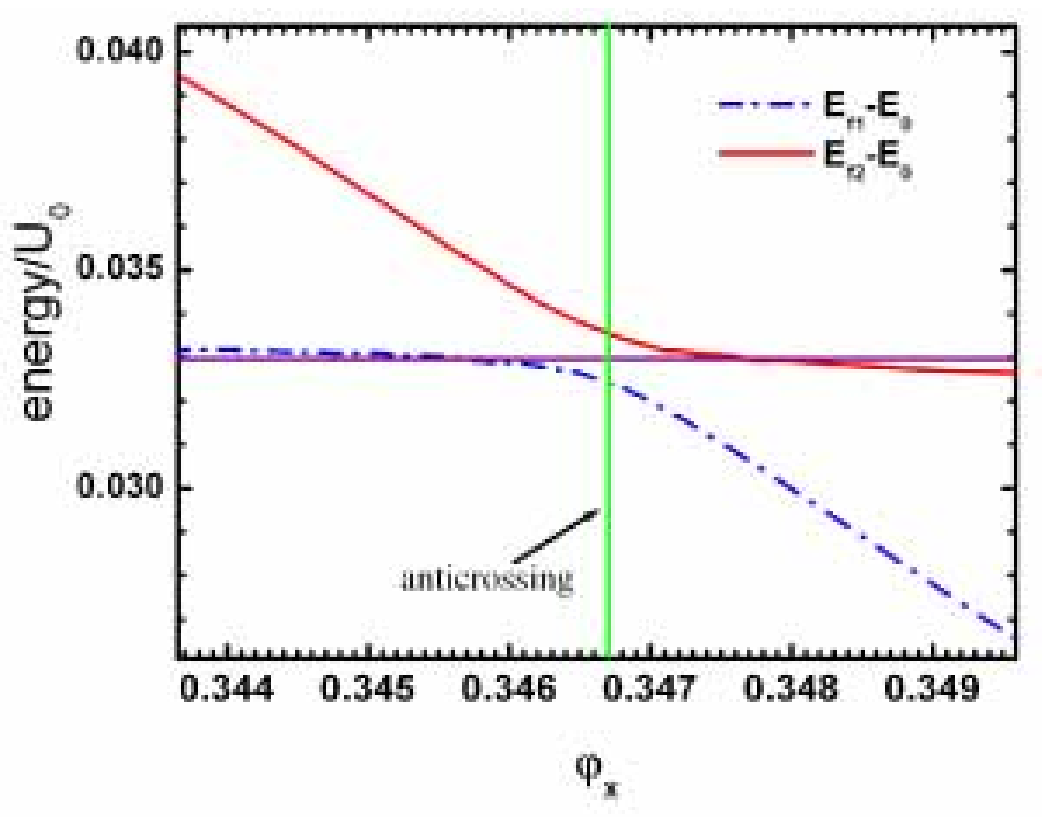}
\caption{\label{anticrossing} Curves $E_{f_1} - E_0$ vs.
$\varphi_x$ and $E_{f_2} - E_0$ vs. $\varphi_x$. All the energies
are referred to the barrier top. The horizontal line is the
external field of frequency $\nu$, it crosses the energy curves
$E_{f_1}$-$E_{0}$ and $E_{f_2}$-$E_{0}$ in two points. The
vertical line represents the coordinate corresponding to the
anticrossing point. Comparison of Figs.\ref{Wtre}
-\ref{anticrossing} shows that the first peak correspond to the
resonant pumping from the ground state in the left well to the
state with energy $E_{f_{1}}$ while the third peak corresponds to
the resonant pumping from the ground state from the ground state
to the state with energy $E_{f_{2}}$. For all the parameters
values are: $\beta_L=1.75$, $C=0.1\,pF$, $L=210\,pH$.}
\end{figure}

\begin{figure}[h]
\includegraphics[bb =   0 0 216 163]{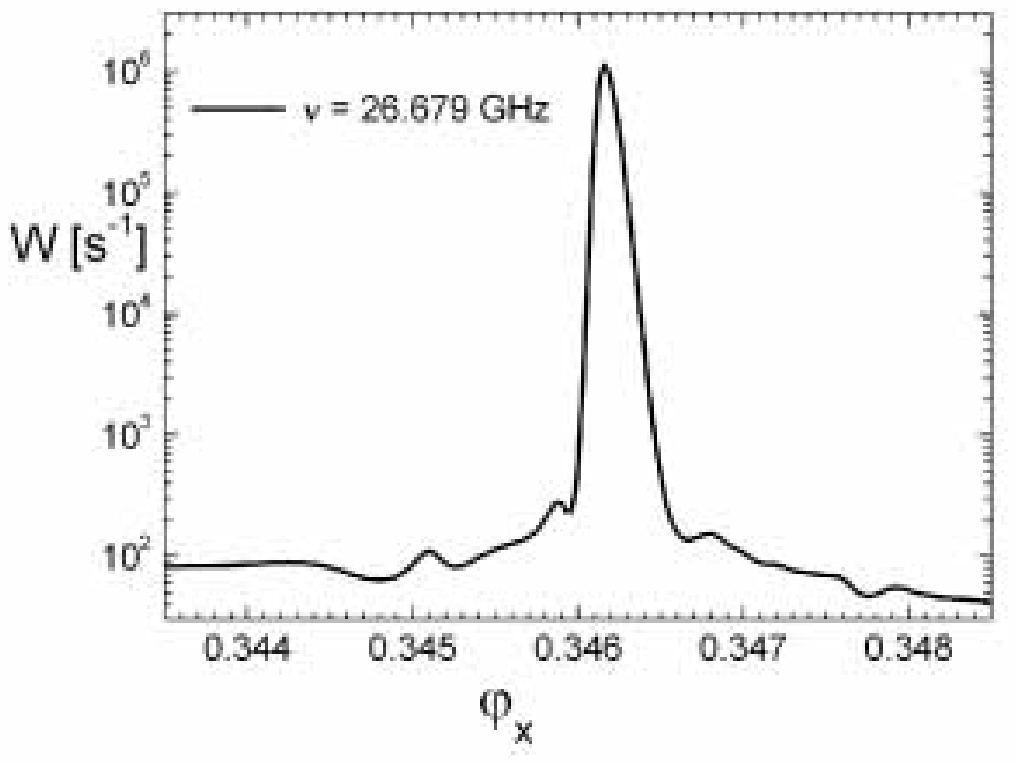}
\caption{\label{Wuno}Transition probability
 W vs. $\varphi_x$. Here the W presents only one peak due to the
resonance tunneling. The plot was obtained by using the following
parameters for the numerical simulation: $\beta_L$=1.75, L=210 pH,
C=0.1 pF, and $R_{eff}$=8 M$\Omega$.}
\end{figure}

\begin{figure}[h]
\includegraphics[bb =  0 0 216 162]{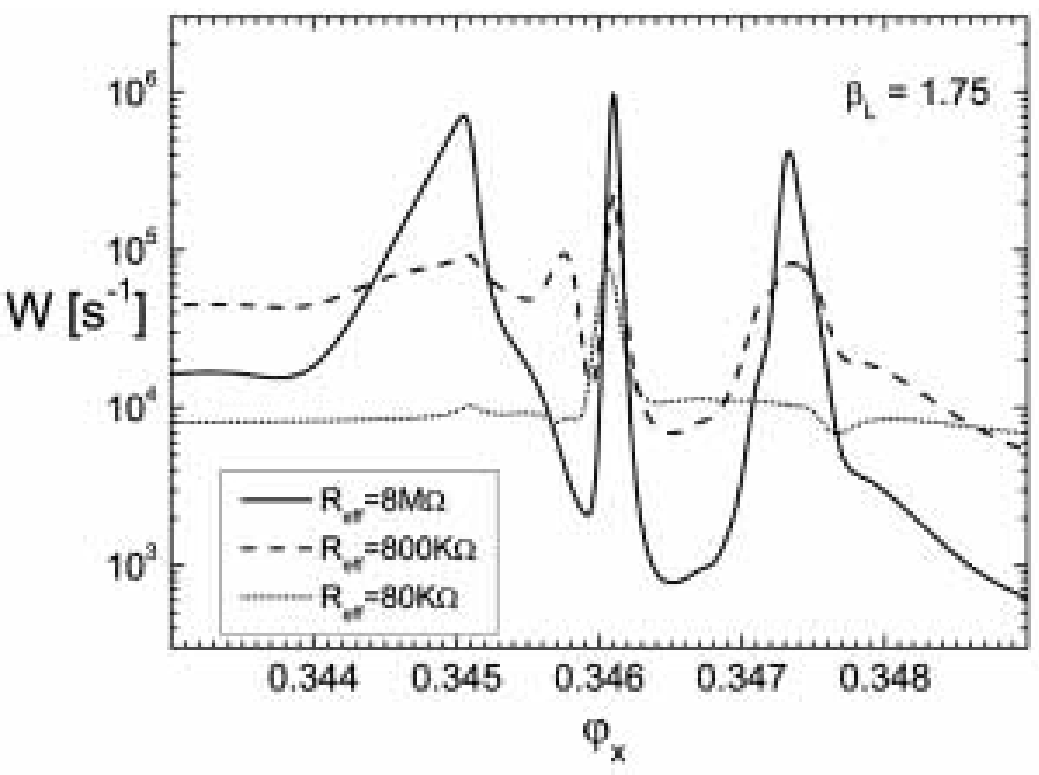}
\caption{\label{WdiR}Transition probability
 W vs. $\varphi_x$ for different values of the effective resistance $R_{eff}$.
  The plot was obtained by using the following
parameters for the numerical simulation: $\beta_L$=1.75, L=210 pH,
C=0.1 pF, $\nu$= 25.756 GHz.}
\end{figure}

\begin{figure}[h]
\includegraphics[bb = 0 0 216 167]{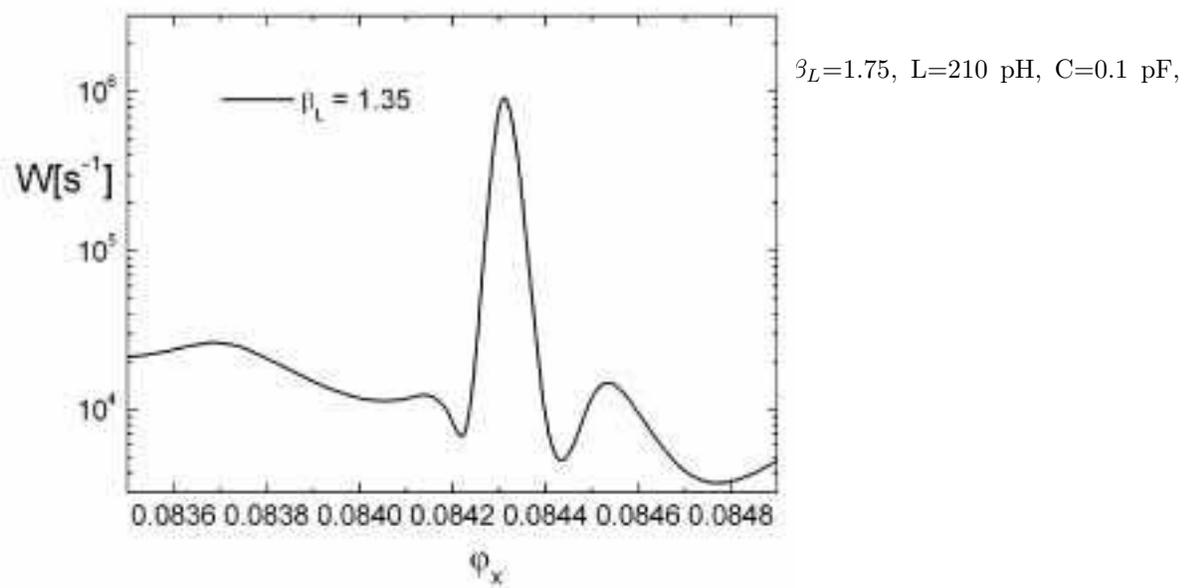}
\caption{\label{Wbetal135}Transition probability
 W vs. $\varphi_x$ for $\beta_L$=1.35, L=162 pH,
C=0.1 pF, $R_{eff}$=8 M$\Omega$ and $\nu$= 24.456 GHz.}
\end{figure}

In this section we suggest a set of SQUID parameters useful in
order to observe experimentally the effect here studied with this
new theoretical approach. The experiment can be realized by
following the experimental scheme described in \cite{ELQ3}. The
external parameter $\varphi_x$ biasing the rf-SQUID is varied as a
function of the time until a switching from one fluxoid state to
another one is observed. The statistical distribution of the
switching value $\varphi_x$ can be measured by repeating the
process many times (about $10^4$ measurements). From the switching
value distribution is straightforward to obtain the escape rate
$W$ and compare data and theory. The first significant comparison
is then the dependence of the escape rate as a function of
$\varphi_x$, showing the characteristic peaks due to ELQ and MQC
phenomena. For these simulations we consider a temperature T=50 mK
and some reasonable rf-SQUID parameters, as: $\beta_L$=1.75, L=210
pH, C=0.1 pF, and $R_{eff}$=8 M$\Omega$.
\\In the extremely underdamped limit, the number of peaks of the
transition probability $W$ vs. $\varphi_x$, can be three or one
depending on the pumping frequency value $\nu$.\\
\\In the three peaks curves (Fig.\ref{Wtre}), the central peak is always connected
with the  resonant tunneling, while the two other peaks are
connected with the resonant pumping. The central peak is due to
the coherent tunneling between levels $f_1$ and $f_2$. As a
consequence, it's fixed in position and height, even if we change
the frequency $\nu$, since it is correlated to the anticrossing
point, that does not depend on the external irradiation.  On the
contrary, the other two peaks have a shift depending on the
frequency. The left side peak corresponds to a transition from the
ground state in the left well to the state $f_1$, while the right
side peak is connected to the transition from the ground state to
the state $f_2$. Of course these transitions occur at values of
the external flux $\varphi_x$ which depend on the pumping
frequency, as shown in Fig.\ref{anticrossing}. Infact, if we
represent the external field of frequency $\nu$ with a horizontal
line, the three peaks curve is possible only when the frequency
crosses both the energy curves $E_{f_1}$-$E_{0}$ and
$E_{f_2}$-$E_{0}$.\\
We stress that we can observe the three peaks curve just in a small
frequency range around the anticrossing point. For all other
frequencies we find the one peak curve (Fig.\ref{Wuno}), where the
only peak is due to the resonant tunneling.
\\For these simulations we have  considered pumping frequencies ranging from
$\nu = \omega/2\pi$ = 25.6 GHz to $\nu$ = 26.8 GHz. These values
assure that the excited pumped level in the left potential well is
close to some level in right
potential well.\\
Moreover it is interesting to study the dependence of the peaks on
the effective dissipation described in the RSJ model by the
effective resistance $R_{eff}$, and this is shown in Fig.\ref{WdiR}.
As expected \cite{RMQTdinopaolo}, by decreasing dissipation, that is
by increasing $R_{eff}$,  peaks resolution is enhanced.
\\
Also the $\beta_{L}$ parameter can be varied in the experiments
\cite{Lukens} and therefore we present some numerical simulations to
study the transition probability W vs.$\varphi_x$ for two different
$\beta_{L}$ values  (see fig.\ref{Wbetal215}, fig.\ref{Wbetal175}
and fig.\ref{Wbetal135}). It's worth noting that to vary $\beta_{L}$
for the rf-SQUID, we vary only the value of the inductance L (see
eq.\ref{definizioni})  and  a variation of $\beta_{L}$ corresponds
to the variation of the height of the potential barrier. We stress
that, when we increase the $\beta_{L}$ value, the transition
probability becomes smaller and we need to increase  the external
parameter to observe the peaks.
\begin{figure}[h]
\includegraphics[bb = 0 0 216 157]{Wbeta215BNL.eps}
\caption{\label{Wbetal215}Transition probability
 W vs. $\varphi_x$ for $\beta_L$=2.15, L=258 pH,
C=0.1 pF, $R_{eff}$=8 M$\Omega$ and $\nu$=32.818 GHz.}
\end{figure}\\
\begin{figure}[h]
\includegraphics[bb = 0 0 216 161]{Wbetal175BNL.eps}
\caption{\label{Wbetal175}Transition probability
 W vs. $\varphi_x$ for $\beta_L$=1.75, L=210 pH,
C=0.1 pF, $R_{eff}$=8 M$\Omega$ and $\nu$= 25.756 GHz.}
\end{figure}\\
\\
Finally we  also made a study of the dependence of the peaks
distribution and behavior on the capacitance C. We observed that
by decreasing the C value, the peaks in the transition probability
become higher and more enhanced. \\The best set of values for
these parameters is when the $\beta_L$
 and the capacitance C values are not too large.
\begin{figure}[h]
\includegraphics[bb = 0 0 216 161]{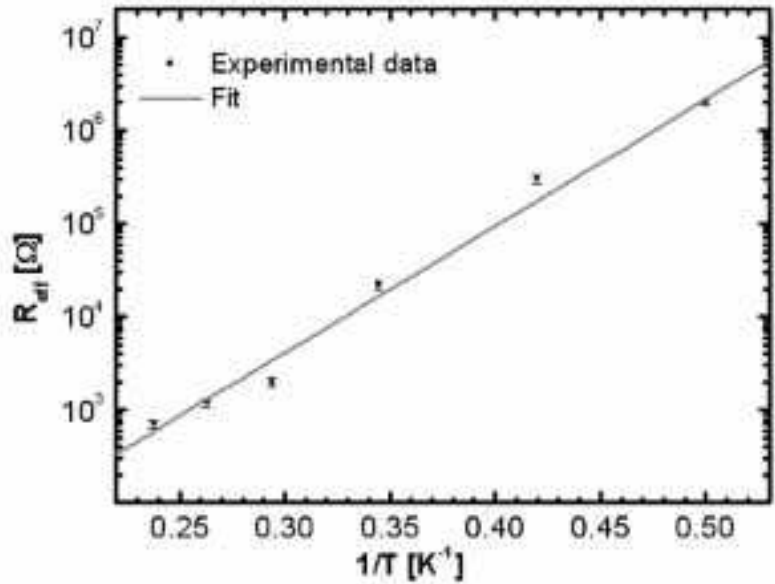}
\caption{\label{RdiT}Dots are experimental data for $R_{eff}$ in our
devices at different temperatures, while the line is the linear fit,
corresponding to the expression
$R_{eff}=R_0\,\exp\left(\Delta/k_{B}T\right)$.}
\end{figure}\\
\\The predicted process can be observed by escape rate
measurements. Since  we want study Macroscopic Quantum Phenomena, we
projected and realized devices with a low dissipation level. The
preliminary measurements on our devices have confirmed a really low
dissipation level \cite{MQT3}, with a really high effective
resistance, as shown in Fig.\ref{RdiT}. These results indicate that
our devices are good candidates for quantum measurements.
\section{Conclusions}
The results presented here describe the small viscosity limit in the
problem of resonant quantum tunneling for an rf-SQUID, whose
parameters can be determined with independent measurements. All the
physical quantities
here considered are very sensitive to the external parameters.\\
It's worth noting that we considered  levels  below the top of the
classical energy barrier. This condition is essential, since in this
case the levels (in the absence of coherence) can  be exclusively
associated with one well and moreover
it's a necessary condition in order to observe tunneling phenomena.\\
The transition probability function W from the left potential well
to the right one has been calculated by varying $\varphi_x$ and for
different values of the pumping frequency $\nu$, of the $\beta_{L}$
parameter and for different values of the effective resistance
$R_{eff}$. The results obtained here suggest that the resonant
quantum tunneling is a convenient tool for
investigating macroscopic quantum coherence phenomena.\\
The dependence of the transition probability W on the external
parameter $\varphi_x$  shows three peaks (Fig.\ref{Wtre}). The first
is connected with the resonant tunneling and the others two are
associated with the resonant pumping toward the two non-localized
energy levels. The relative positions of these two peaks strictly
depend on the pumping frequency and on the external parameter
biasing the rf-SQUID. Moreover the transition probability W  is
obtained as a function of the potential parameters and of the shunt
resistance $R_{eff}$.
\\Preliminary experimental measurements on our devices confirmed a
low dissipation level \cite{MQT3}, with a high effective resistance,
as shown in Fig.\ref{RdiT}. These results, albeit still unconfirmed
by a wider and more accurate experimental campaign, would suggest
potential for interesting observations regarding the resonant
phenomena in the presence of an external microwave irradiation of
proper frequency.
\newpage
\appendix
\section{}
From eqs.(\ref{densmat1}), (\ref{densmat2}) we obtain two equations
depending on the quantities ${\cal D}_1$ and ${\cal F}_1$
\begin{eqnarray}\label{eqF1}
\fl {\cal
F}_1\left[\left(\frac{E_{f_2}-E_{f_1}}{\hbar}\right)-2i\gamma_1
\right]+iW_{f_1\,f_2}^{f_1\,f_2}{\cal D}_1= \frac{i{\cal
I}^2}{16e^2}\cdot\left[\frac{\langle 0\vert\varphi\vert f_1\rangle
\langle f_2\vert\varphi\vert 0\rangle
W_{f_1\,f_2}^{0\,\,0}}{\left(\omega-\left(\frac{E_{f_2}-E_{0}}{\hbar}\right)\right)^2-\gamma_1\gamma_2
+W_{f_1\,f_2}^{0\,\,0}W_{f_2\,f_1}^{0\,\,0}+i(\gamma_1+\gamma_2)\left(\omega-\left(\frac{E_{f_2}-E_{0}}{\hbar}\right)\right)}\right]
\\
\fl iW_{f_2\,f_1}^{f_2\,f_1}{\cal F}_1+{\cal
D}_1\left[\left(\frac{E_{f_2}-E_{f_1}}{\hbar}\right)-2i\gamma_2\right]=
\frac{i{\cal I}^2}{16e^2}\cdot\left[\frac{\langle 0\vert\varphi\vert
f_1\rangle \langle f_2\vert\varphi\vert 0\rangle
W_{f_2\,f_1}^{0\,\,0}}{\left(\omega-\left(\frac{E_{f_1}-E_{0}}{\hbar}\right)\right)^2-\gamma_1\gamma_2+W_{f_1\,f_2}^{0\,\,0}W_{f_2\,f_1}^{0\,\,0}-i(\gamma_1+\gamma_2)\left(\omega-\left(\frac{E_{f_1}-E_{0}}{\hbar}\right)
\right)}\right]
\end{eqnarray}
Solution of equations (\ref{eqF1}) is
\begin{eqnarray}\label{eqF1_sol}
\fl {\cal F}_1=\frac{i{\cal I}^2}{16e^2}\cdot\frac{\langle
0\vert\varphi\vert f_1\rangle \langle f_2\vert\varphi\vert
0\rangle}{\left(\frac{E_{f_2}-E_{f_1}}{\hbar}\right)^2-4\gamma_1
\gamma_2
-2i(\gamma_1+\gamma_2)\left(\frac{E_{f_2}-E_{f_1}}{\hbar}\right)+
W_{f_1\,f_2}^{f_1\,f_2}W_{f_2\,f_1}^{f_2\,f_1}}\cdot\nonumber\\
\fl \cdot\Bigg[\frac{\left[\frac{E_{f_2}-E_{f_1}}{\hbar}-2i\gamma_2
\right]W_{f_1\,f_2}^{0\,\,0}}{\left(\omega-\frac{E_{f_2}-E_{0}}{\hbar}\right)^2-\gamma_1\gamma_2
+W_{f_1\,f_2}^{0\,\,0}W_{f_2\,f_1}^{0\,\,0}+i(\gamma_1+\gamma_2)\left(\omega-\frac{E_{f_2}-E_{0}}{\hbar}\right)}\\
\fl
-i\frac{W_{f_1\,f_2}^{f_1\,f_2}W_{f_2\,f_1}^{0\,\,0}}{\left(\omega-\frac{E_{f_1}-E_{0}}{\hbar}\right)^2-\gamma_1\gamma_2
+W_{f_1\,f_2}^{0\,\,0}W_{f_2\,f_1}^{0\,\,0}-i(\gamma_1+\gamma_2)\left(\omega-\frac{E_{f_1}-E_{0}}{\hbar}\right)}\Bigg]\nonumber
\\
\fl {\cal D}_1=\frac{i{\cal I}^2}{16e^2}\cdot\frac{\langle
0\vert\varphi\vert f_1\rangle \langle f_2\vert\varphi\vert
0\rangle}{\left(\frac{E_{f_2}-E_{f_1}}{\hbar}\right)^2-4\gamma_1
\gamma_2 -2i(\gamma_1+\gamma_2)\frac{E_{f_2}-E_{f_1}}{\hbar}+
W_{f_1\,f_2}^{f_1\,f_2}W_{f_2\,f_1}^{f_2\,f_1}}\cdot\nonumber\\
\fl \Bigg[\frac{\left[\frac{E_{f_2}-E_{f_1}}{\hbar}-2i\gamma_1
\right]W_{f_2\,f_1}^{0\,\,0}}{\left(\omega-\frac{E_{f_1}-E_{0}}{\hbar}\right)^2-\gamma_1\gamma_2
+W_{f_1\,f_2}^{0\,\,0}W_{f_2\,f_1}^{0\,\,0}-i(\gamma_1+\gamma_2)\left(\omega-\frac{E_{f_1}-E_{0}}{\hbar}\right)}\\
\fl
-i\frac{W_{f_2\,f_1}^{f_2\,f_1}W_{f_1\,f_2}^{0\,\,0}}{\left(\omega-\frac{E_{f_2}-E_{0}}{\hbar}\right)^2-\gamma_1\gamma_2
+W_{f_1\,f_2}^{0\,\,0}W_{f_2\,f_1}^{0\,\,0}+i(\gamma_1+\gamma_2)\left(\omega-\frac{E_{f_2}-E_{0}}{\hbar}\right)}\Bigg]\nonumber
\end{eqnarray}
The explicit expression for the transition matrix elements
$W_{f_1\,f_2}^{f_1\,f_2}$ and $W_{f_1\,f_2}^{0\,\,0}$ is given below
\begin{eqnarray}\label{defW}
\fl
W_{f_1\,f_2}^{f_1\,f_2}=\frac{\pi}{R_{eff}e^2}\left(1+\tanh\left(\frac{E_{f_2}-E_{f_1}}{2k_{B}T}\right)\right)\frac{E_{f_2}-E_{f_1}}{\pi}
\coth\left(\frac{E_{f_2}-E_{f_1}}{2k_{B}T}\right)\vert\langle
f_1|\exp^{i\frac{\varphi}{2}}\vert f_2\rangle\vert ^2\\
\fl
W_{f_1\,f_2}^{0\,\,0}=\frac{\pi}{2R_{eff}e^2}\left(1+\tanh\left(\frac{E_{f_2}-E_{f_1}}{4k_{B}T}\right)\right)\frac{E_{f_2}-E_{f_1}}{2\pi}
\coth\left(\frac{E_{f_2}-E_{f_1}}{4k_{B}T}\right)\cdot\nonumber\\
\fl \cdot\left[\langle 0|\exp^{i\frac{\varphi}{2}}\vert 0\rangle
\langle f_1|\exp^{-i\frac{\varphi}{2}}\vert f_2\rangle+\langle
0|\exp^{-i\frac{\varphi}{2}}\vert 0\rangle \langle
f_1|\exp^{i\frac{\varphi}{2}}\vert f_2\rangle \right]\\
\fl
W_{f_2\,f_1}^{0\,\,0}=\exp\left(-\frac{E_{f_2}-E_{f_1}}{2k_{B}T}\right)W_{f_1\,f_2}^{0\,\,0}
\end{eqnarray}
\newpage
\section{}
For the potential, given by eq.(\ref{potSQUID}) we have
\begin{eqnarray*}\label{A1}
\fl \varphi_{top}=\varphi_x + \beta_L\sin (\varphi_{top}) \\
\fl
\frac{\partial\varphi_{top}}{\partial\varphi_{x}}=-\frac{1}{\beta_L\cos
(\varphi_{top})-1}\nonumber\\
\fl \frac{\partial U}{\partial\varphi_{x}}=-U_0(\varphi-\varphi_{x})
\end{eqnarray*}
\begin{eqnarray*}\label{A1bis}
\fl U_1=\frac{U_0}{2}\left(\beta_L\cos (\varphi_{top})-1\right)\\
\fl \frac{\partial U_1}{\partial\varphi_{x}}=\frac{1}{2}\beta_L
U_0\frac{\sin (\varphi_{top})}{\beta_L\cos (\varphi_{top})-1}
\end{eqnarray*}
\begin{eqnarray*}\label{A1tris}
\fl U_{top}=U_0\left[\frac{1}{2}(\varphi_{top}-\varphi_{x})^2+\beta_L\cos\varphi_{top}\right]\\
\fl \frac{\partial
U_{top}}{\partial\varphi_x}=U_0\left[\frac{\beta_L\sin\varphi_{top}-(\varphi_{top}-\varphi_{x})}{\beta_L\cos
(\varphi_{top})-1}-(\varphi_{top}-\varphi_{x}) \right]
\end{eqnarray*}
From equations (\ref{def_U}), (\ref{defA_B})  we obtain all other
derivatives that should be found for calculation of the quantities
$\alpha_{1,2}$, $\beta_{1,2}$.
\begin{eqnarray}\label{A2}
\fl \frac{\partial\lambda}{\partial E}=-\sqrt{\frac{2M}{U_1}}\\
\fl \frac{\partial\lambda}{\partial\varphi _{x}}=U_0\sqrt{\frac{2M}{U_1}}\left[\frac{1}{\beta_{L}\cos(\varphi_{top})-1}\left[\beta_{L}\sin\varphi_{top}\left(1-\frac{U_{top}-E}{4U_1}\right)-(\varphi_{top}-\varphi_{x})\right]-(\varphi_{top}-\varphi_{x})\right]\\
\fl
\frac{\partial\chi}{\partial\lambda}=\frac{1}{2}\psi\left(\frac{1}{2}\right)+\lambda^2\sum_{k=0}^{\infty}{\frac{1}{(2k+1)((2k+1)^2+\lambda^2)}}
\end{eqnarray}
Now by using equations (\ref{defPhi_1}), (\ref{defPhi_2}),
(\ref{A1}), (\ref{A2}) we obtain the value of all coefficients
$\alpha_{1}$, $\alpha_{2}$, $\beta_{1}$, $\beta_{2}$.
\begin{eqnarray}\label{beta1}
\fl \beta_1=\frac{\partial\Phi_1}{\partial
E}+\frac{1}{2}\frac{\partial\chi}{\partial\lambda}\frac{\partial\lambda}{\partial
E}=
\sqrt{\frac{M}{2}}\int\limits_{\tilde{\varphi_1}}^{\varphi_{top}}{d\varphi}\left(\frac{1}{\sqrt{U_{top}-U(\varphi)}}
-\frac{\sqrt{\varphi_{top}-\tilde{\varphi_1}}}{(\varphi_{top}-\varphi)\sqrt{U_1(\varphi-\tilde{\varphi_1})}}\right)\\
\fl
+\sqrt{\frac{M}{2U_1}}\ln\left(\frac{8(MU_1)^{1/4}(\varphi_{top}-\tilde{\varphi_1})}{2^{1/4}}\right)-\sqrt{\frac{M}{2U_1}}\frac{\partial\chi
}{\partial\lambda}\nonumber
\end{eqnarray}
\begin{eqnarray}\label{beta2}
\fl \beta_2=\frac{\partial\Phi_2}{\partial
E}+\frac{1}{2}\frac{\partial\chi}{\partial\lambda}\frac{\partial\lambda}{\partial
E}=
\sqrt{\frac{M}{2}}\int\limits^{\tilde{\varphi_4}}_{\varphi_{top}}{d\varphi}\left(\frac{1}{\sqrt{U_{top}-U(\varphi)}}
-\frac{\sqrt{\tilde{\varphi_4}-\varphi_{top}}}{(\varphi-\varphi_{top})\sqrt{U_1(\tilde{\varphi_4}-\varphi)}}\right)\\
\fl
+\sqrt{\frac{M}{2U_1}}\ln\left(\frac{8(MU_1)^{1/4}(\tilde{\varphi_4}-\varphi_{top})}{2^{1/4}}\right)-\sqrt{\frac{M}{2U_1}}\frac{\partial\chi
}{\partial\lambda}\nonumber
\end{eqnarray}
\begin{eqnarray}\label{alpha1}
\fl
\alpha_1=\frac{\partial\Phi_1}{\partial\varphi_{x}}+\frac{1}{2}\frac{\partial\chi}{\partial\lambda}\frac{\partial\lambda}{\partial
\varphi_x}=U_0\sqrt{\frac{M}{2}}\int\limits_{\varphi_1}^{\varphi_2}{d\varphi\frac{\varphi-\varphi_{x}}{\sqrt{E-U(\varphi)}}+\left(\frac{1}{2}\frac{\partial\chi}{\partial\lambda}+\frac{1}{4}\ln\left(
\frac{2}{\lambda}\right)\right)U_0\sqrt{\frac{2M}{U_1}}}\cdot\\
\fl
\cdot\left[\frac{1}{\beta_{L}\cos\varphi_{top}-1}\left[\beta_{L}\sin\varphi_{top}\left(1-\frac{U_{top}-E}{4U_{1}}\right)-(\varphi_{top}-\varphi_{x})\right]-(\varphi_{top}-\varphi_{x})\right]\nonumber
\end{eqnarray}
\begin{eqnarray}\label{alpha2}
\fl \alpha_2=\frac{\partial\Phi_2}{\partial
\varphi_x}+\frac{1}{2}\frac{\partial\chi}{\partial\lambda}\frac{\partial\lambda}{\partial
\varphi_x}=
U_0\sqrt{\frac{M}{2}}\int\limits_{\varphi_3}^{\varphi_4}{d\varphi\frac{\varphi-\varphi_{x}}{\sqrt{E-U(\varphi)}}+\left(\frac{1}{2}\frac{\partial\chi}{\partial\lambda}+\frac{1}{4}\ln\left(
\frac{2}{\lambda}\right)
\right)U_0\sqrt{\frac{2M}{U_1}}}\cdot\\
\fl \cdot\left[\frac{1}{\beta_L\cos\varphi_{top}-1} \left[
\beta_L\sin\varphi_{top}\left(1-\frac{U_{top}-E}{4U_1}\right)-(\varphi_{top}-\varphi_{x})\right]-(\varphi_{top}-\varphi_{x})\right]\nonumber
\end{eqnarray}
\newpage
\section{} Transition matrix elements between states
close to the barrier top can be calculated in quasiclassical
approximation if the parameter $\lambda$ (see eq.(\ref{f_onda})) is
$\lambda\geq 1$. Consider first the matrix element $\langle L\vert\zeta\vert f_{1}\rangle$. With the help of eqs.(\ref{f_onda}), (\ref{def_psiL}), (\ref{def_psiR}) we obtain\\
\begin{equation}\label{uno}
\fl \langle L\vert\zeta\vert
f_{1}\rangle=\frac{1}{G_{L}G_{E_{f_1}}}\int\limits_{\varphi_1({E_{f_1}})}^{\varphi_{2}({E_{f_1}})}{d\varphi\frac{\sin\left(
\pi/4+\int\limits_{\varphi_1({E_{L}})}^{\varphi}{d\varphi\sqrt{\left(2M\left(E_{L}-U(\varphi)\right)\right)}}\right)}{\left(2M\left(E_{L}-U(\varphi)\right)\right)^{1/4}\left(2M\left(E_{f_1}-U(\varphi)\right)\right)^{1/4}}\,\zeta
\sin\left(
\pi/4+\int\limits_{\varphi_1({E_{f_1}})}^{\varphi}{d\varphi\sqrt{\left(2M\left(E_{f_{1}}-U(\varphi)\right)\right)}}\right)}
\end{equation}
that is equal to
\begin{equation}
\label{due} \fl \langle L\vert\zeta\vert
f_{1}\rangle=\frac{1}{2G_{L}G_{E_{f_1}}}
\int\frac{d\varphi}{\sqrt{\left(2M\left(E_{f_{1}L}-U(\varphi)\right)\right)}}\,\zeta\cos\left(\left(E_{f_{1}}-E_{L}\right)
\int_{\varphi_{1}(E_{f_{1}L})}^{\varphi}\frac{Md\varphi}{\sqrt{2M\left(E_{f_{1}L}-U(\varphi)\right)}}\right)
\end{equation}
To calculate the last integral we can use the time variable $t$
according to the classical equation of motion
\begin{equation}\label{motion}
\fl \frac{M}{2}\left(\frac{\partial\varphi}{\partial t} \right)^2 =
E-U(\varphi)
\end{equation}
As a consequence,  eq.(\ref{due}) becomes
\begin{equation}
\label{tre}\fl  \langle L\vert\zeta\vert f_{1}\rangle =
\frac{1}{2MG_{L}G_{E_{f_1}}}\int\limits_{0}^{T_{1}/2}{dt\zeta
\cos\left(\frac{2\pi}{T_1}t\ell \right)}
\end{equation}
where $T_1$ is the period of the classical motion in the left
potential well with energy E and $\ell$ is the  number of states
between energy levels $E_{f_1}$ and $E_L$ plus one. In our case
$\ell=1$. The initial value for $\varphi$ is $\varphi
(0)=\varphi_1$.
\begin{equation}\label{eqT1}
\fl
T_1=2M\int\limits_{\varphi_3}^{\varphi_4}{\frac{d\varphi}{\sqrt{\left(2M\left(E-U(\varphi)\right)\right)}}}
\end{equation}
In the same way we obtain transition the matrix element $\langle
R\vert\zeta\vert f_{1}\rangle$.
\newline\\In eq.(\ref{uno}) $\varphi_1({E_{f_1}})$ is the first
crossing point of the energy level $E_{f_1}$ in the left well and
$\varphi_2({E_{f_1}})$ is the second crossing point of the energy
level $E_{f_1}$ in the left well. \\In the same way in
eq.(\ref{due}) $\varphi_{1}(E_{f_{1}L})$ is the first crossing
point of the energy level $E_{f_{1}L}$ in the left well, where
$E_{f_{1}L}$ is defined as
\begin{equation}\label{Ef1L}
\fl E_{f_{1}L}=\frac{E_{f_1} +E_{L}}{2}
\end{equation}
Consider now the transition matrix element $\langle f_2\vert\
\zeta\vert f_1\rangle$. From eq.(\ref{f_onda}) we find
\begin{equation}\label{f2zetaf1A}
\fl \langle f_2\vert\ \zeta\vert
f_1\rangle=\frac{\hbar}{2G_{f_1}G_{f_2}}\left[\int\limits_{\varphi_1}^{\varphi_2}{d\varphi\frac{\zeta}{\sqrt{\left(2M\left(E-U(\varphi)\right)\right)}}}
+B_{f_1}B_{f_2}\int\limits_{\varphi_3}^{\varphi_4}{d\varphi\frac{\zeta}{\sqrt{\left(2M\left(E-U(\varphi)\right)\right)}}}
\right]
\end{equation}
where $\varphi_{1,2,3,4}$ are the ``turning points" and $E$ here is
$E=\frac{\left(E_{f_1}+E_{f_2}\right)}{2}$.
\\By using eq.(\ref{motion}) we reduce the expression (\ref{f2zetaf1A}) to the
form{\small
\begin{equation}\label{f2zetaf1B}
\fl \langle f_2\vert\ \zeta\vert
f_1\rangle=\frac{\hbar}{2MG_{f_1}G_{f_2}}\left[\int\limits_{0}^{T_{1}/2}{dt\zeta}+B_{f_1}B_{f_2}\int\limits_{0}^{T_{2}/2}{dt\zeta}\right]
\end{equation}}
In eq.(\ref{f2zetaf1B}) the first integral is taken over the left
potential well and the second one over the right potential well and
$T_{1}$,
$T_{2}$ are the periods of the classical motion in the left and in the right well respectively. \\
To improve eq.(\ref{f2zetaf1B}) we should take into account
ortogonality of wavefunctions relative to the energies $E_{f_1}$,
$E_{f_2}$ . As result we obtain{\footnotesize
\begin{equation}\label{f2zetaf1C}
\fl \langle f_2\vert\ \zeta\vert
f_1\rangle=\hbar\frac{1-B_{f_1}B_{f_2}}{2MG_{f_1}G_{f_2}}\left[\frac{T_2}{T_1+T_2}\int\limits_{0}^{T_{1}/2}{dt\zeta}-\frac{T_1}{T_1+T_2}\int\limits_{0}^{T_{2}/2}{dt\zeta}\right]
\end{equation}}
\ack{The research of one of us (Yu.N.O.) is supported by Cariplo
Foundation-Italy, CRDF USA under Grant No. RP1-2565-MO-03 and the
Russian Foundation of Basic Research. This work has been partially
supported by MIUR under Project ``Reti di Giunzioni Josephson per la
Computazione e l'Informazione Quantistica-JOSNET".}
\section*{References}
\bibliographystyle{unsrt}
\bibliography{art14ottobre}% Produces the bibliography via BibTeX.
\end{document}